# The Realizations of Steganography in Encrypted Domain

Yan Ke[0000-0002-6229-9998] Minqing Zhang[0000-0002-8783-3608] Jia Liu[0000-0001-8104-0079]
Xiaoyuan Yang

[1] Engineering University of PAP, Xi'an, 71086, China
[2] Key Laboratory of PAP for Cryptography and Information Security, Xi'an, 71086, China
[3] School of Cryptography Engineering in Engineering University of PAP
15114873390@163.com

**Abstract.** With the popularization and application of privacy protection technologies in cloud service and social network, ciphertext has been gradually becoming a common platform for public to exchange data. Under the cover of such a platform, we propose steganography in encrypted domain (SIED) in this paper to realize a novel method to realize secret communication Based on Simmons' model of prisoners' problems, we discuss the application scenarios of SIED. According to the different accesses to the encryption key and decryption key for secret message sender or receiver, the application modes of SIED are classified into four modes. To analyze the security requirments of SIED, four levels of steganalysis attacks are introduced based on the prior knowledge about the steganography system that the attacker is assumed to obtain in advance. Four levels of security standards of SIED are defined correspondingly. Based on the existing reversible data hiding techniques, we give four schemes of SIED as practical instances with different security levels. By analyzing the embedding and extraction characteristics of each instance, their SIED modes, application frameworks and security levels are discussed in detail.

## 1 Introduction

The modern secret communication technology originated from the military demands of the secret communication since World War II, and the realization of secret communication mainly consists of two major technologies: cryptography and steganography [1]. Steganography is an important covert communication technique, the characteristic of which is to maintain the undetectability of the existence of the secret communication. The realization of steganography technology is derived from the various public communication platforms with popular applications in social lives, such as website pages, social networks, images or video shows, etc.

The prototype of steganography was first defined with modern terminology in Simmons' founding work on subliminal channels and the prisoners' problems [2]. In prisoner's problems [2], Alice and Bob are in jail and they want to devise an escape plan by exchanging hidden messages in innocent-looking covers (*e.g.*, natural images). These covers are conveyed to one another by a common warden, Eve, who can eavesdrop all covers and can choose to interrupt the communication if they appear to be a stego-cover. In this model, the secret communication consists of three basic elements: 1. secret message, 2. cover, and 3. open channel. The open channel of existing steganography schemes is mainly derived from the public communication platforms. Communication platforms have been enriched with the development of the social environments. Since privacy protection techniques are widely introduced in cloud services, blockchain technology, and federated learning. Ciphertext are gradually becoming a common platform for the public. How to realize covert communication under the cover of ciphertext environments is drawing attentions.

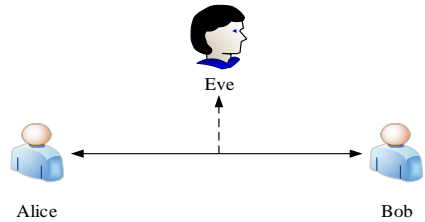

**Fig. 1.** The prisoners' problems

This paper mainly focuses on the characteristics and security requirements of the steganography under the condition of communication in encrypted domain, and introduces several feasible realizations of steganography in encrypted domain (SIED), and analyzes the principles and security requirements of SIED.

## 2    Application modes of SIED

### 2.1    Common Applications of Ciphertext

Combining the prisoner problem to illustrate the differences between SIED and traditional steganography, SIED are realized on the cryptosystem, in which Alice and Bob must perform encryption operations first. Then Alice and Bob could embed (extract) secret messages into (out of) the ciphertext. The data transmitted through Eve is the ciphertext carrying additional secret messages, *i.e.*, stego-ciphertext. The proposal of SIED is based on the wide applications of ciphertext in common social life. There exists applications in real life for Alice to exchange messages by ciphertext.

For examples, to protect the user's privacy, existing social networks, cloud services, blockchain and streaming media platforms supports client Alice to encrypt their data before uploading to the service or another client. The data of Alice is usually some information from her daily life, not confidential information. Encryption port is just a common service open access to the public. For some individuals in special environments, whose available communication tools are mainly encrypted channels or cryptosystems, *e.g.*, staff from the core research institution, inter-country manager of enterprises, or security departments, etc. All above could provide a promising tool to cover the communication of SIED.

## 2.2 Applications Modes

In a cryptosystem, let the plaintext be denoted as $P$, the ciphertext as $C$, the encryption key as $K_{Enc}$, the decryption key as $K_{Dec}$, the encryption function as Enc( . , . ), the decryption function Dec(. , .). The relationships among them are as shonw in Eqs. (1)-(2):

$$C = \text{Enc}(K_{Enc}, P) \tag{1}$$

$$P = \text{Dec}(K_{Dec}, C) \tag{2}$$

- Ciphertext $C$ is the cover used to carry secret messages and conceal the existence of the secret communication.

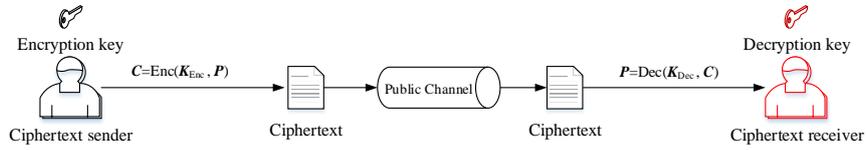

**Fig. 2.** Communication in cryptosystem

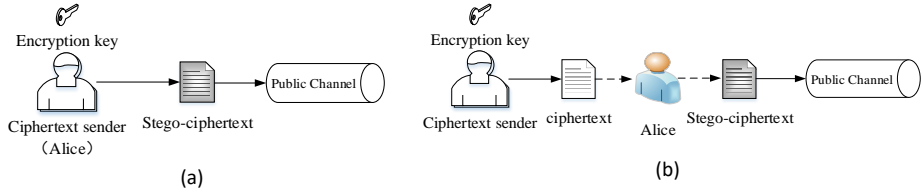

**Fig. 3.** Embedding modes, *i.e.*, the two cases of generating stego-ciphertext by Alice: (a) with encryption key; (b) without encryption key.

- Stego-ciphertext, denoted as C', is the cover that has been embedded with the secret message.
- Secret message, denoted as $m$, is the message to be embedded into the cover for communication.

Let the key for embedding be denoted as $K_{emb}$, the key for secret message extraction as $K_{ext}$, the embedding function as Emb(. , .), the extraction function be denoted as Ext(. , .). The relationships among them are as shonw in Eqs. (3)-(4):

$$C' = \text{Emb}(K_{emb}, C, m) \tag{3}$$

$$m = \text{Ext}(K_{ext}, C') \tag{4}$$

Before introducing the application scenarios of SIED, we first introduce the main participants of the cryptosystem. As shown in Fig. 2, the sender and the receiver have been assigned with encryption key and decryption key in advance for communication. The sender uses the encryption key to encrypt the plaintext to obtain the ciphertext and sends the ciphertext to the receiver through a public (untrusted) channel, and the receiver uses the pre-assigned decryption key to encrypt the ciphertext into plaintext to be able to read the communication content. The SIED is implemented based on the above communication process.

There are two cases for data embedding modes of SIED. The 1$^{st}$ case is that the data hider (Alice) is the sender of the cryptosystem, that is, Alice has access to the encryption key, as shown in Fig. 3(a); the 2$^{nd}$ case is that Alice can only implement embedding after encryption, as shown in Fig. 3(b). Alice has to intercept the original ciphertext of the cryptosystem first, and then replay the embedded ciphertext into the channel and send it to the receiver. In this case, the computational complexity of Alice is relatively high. Considering the characteristics of the modern public key cryptosystem, the encryption key is usually open to the public. Therefore, the 2$^{nd}$ case may be more practical in reality.

There are also two cases for the extraction modes of SIED based on whether encryption key is necessary for data extraction, as shown in Fig. 4. The 1$^{st}$ case is that Bob is a legitimate receiver of the cryptosystem with the decryption key. Bob could extract the secret messages from the stego-ciphertext by using the decryption key. Even if Bob can master the decryption key, he may choose not to use that key for information extraction, depending on the requirements of the extraction function of the SIED algorithm. But in practical applications, the receiver or owner of the ciphertext is usually not the only one, Bob is just one of them. The 2$^{nd}$ case is that Bob is not a legitimate receiver of the cryptosystem. Bob needs to intercept the stego-ciphertext by eavesdropping, and extracts the secret messages from it without using the decryption key.

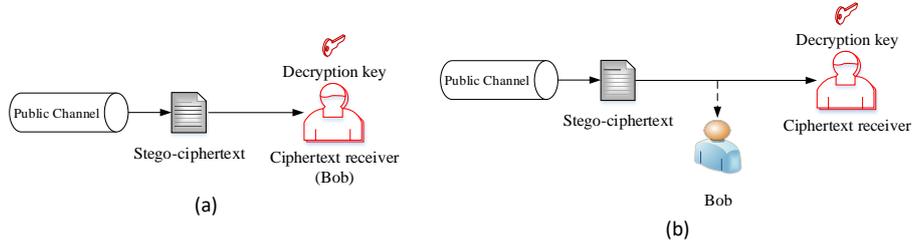

**Fig. 4.** Extraction modes, *i.e.*, the two cases of extracting data from stego-ciphertext by Bob: (a) with decryption key; (b) without decryption key.

**Table 1.** The classification of SIED

| Modes | Alice | Bob | Flowchart |
| --- | --- | --- | --- |
| AC | With encryption key | With decryption key | Fig.3 (a) & Fig. 4(a) |
| EXC | Without encryption key | With decryption key | Fig.3 (b) & Fig. 4(a) |
| EMC | With encryption key | Without decryption key | Fig.3 (a) & Fig. 4(b) |
| AF | Without encryption key | Without decryption key | Fig.3 (b) & Fig. 4(b) |

To sum up, there are four application modes of SIED that can be classified from the perspective of whether Alice or Bob can use the encryption key and decryption key. As shown in Table 1, the application modes include:

- All controlled (AC) mode.

- Extraction controlled (EXC) mode.
- Embedding controlled (EMC) mode.
- All free (AF) mode.

In AF mode, the steganography system is free from the security constraints of cryptosystems and could be applied to more complex applications.

The security of steganography is mainly reflected by the degree of concealing the existence of secret communication. Specifically, it requires the indistinguishability of the stego-cover and the original cover. Below, we specifically analyze the security requirements of SIED.

## 3 Security of SIED

### 3.1 Indistinguishability

The indistinguishability of cover and stego-cover is an important guarantee of stego-security. For the attackers, it is a variety of covers that they can directly obtain in the open channels. Current steganography methods also focus on the indistinguishability, on which steganographic researchers have done a lot of work in recent years [1]. Firstly, the concept of an abstract distance D (D≥0) is given, which quantifies the differentiability between different samples. $D(P(C), P(C'))=0$ indicates that the ciphertext $C$ is indistinguishable from the stego-ciphertext $C'$, which means not only the imperceptibility in the human visual system, but also the undetectability in the sense of statistical analysis [1].

$C$ and $C'$ are indistinguishable iff (if and only if)

$$D(P(C), P(C')) = 0 \tag{5}$$

$C$ and $C'$ are distinguishable iff

$$D(P(C), P(C')) > 0 \tag{6}$$

In practice, it is usually not directly evaluated by D=0. Here, $(t, \varepsilon)$-*indistinguishable* is defined: $C$ and $C'$ are $(t, \varepsilon)$-indistinguishable iff

$$D(P(C), P(C')) < \varepsilon \tag{7}$$

in polynomial time $t$, in which $\varepsilon$ is an arbitrary small value.

There are many specific mathematical tools to indicate the imperceptibility. ITU-T has defined several ways of subjective evaluation criteria, such as ITU-T Rec. P.910. Mean opinion score is the most representative subjective evaluation method for image quality [3]. For the statistical analysis, Cachin used Kullback-Leibler (KL) divergence distance [4] to describe the distribution deviation in steganography system [5], [6]. In [7], F. Cayre first introduced conditional probability to analyze the watermarking secu-rity which was referred to as stego-security [8], where $P(C'/K)$ denotes the model of $C'$ by data hiding key $K$.

$$P(C'/K)) = P(C) \tag{8}$$

The features of models of multimedia that acted as covers in traditional steganography have the characteristics of high dimensionality and high complexity. With the continuous increase of various types of models, the model dimension continues to increase. But the usability of the model still cannot reach the provable security [1]. On the other hand, steganalysis tools based on deep learning or reinforcement learning have made breakthroughs, threaten the practicability of steganography schemes. Fortunately, ciphertext has a natural advantage in guaranteeing theoretical security of steganography, because the ideal distribution model of ciphertext has been given already. According to "Communication theory of secrecy systems" by Shannon in 1949, the standard encrypted data follows randomly uniform distribution. Let the encrypted domain be a finite field $F_q$, the uniform distribution in $F_q$ is denoted as $U(0, q)$. The information entropy of the data sampled from $U(0, q)$ tends to be the largest. Therefore, when considering the embedding function and the constraints of the distribution of the stego-ciphertext, it is obvious that the stego-ciphertext should follow the randomly uniform distribution.

$$P(C') = P(C) = U(0, q) \tag{9}$$

In SIED, Eq. (9) is required to be satisfied after embedding. However, steganography security has not been achieved yet, because Eve here is just assumed to be an eavesdropper capable of stego-cover only attack (SCOA). In practice, what Eve can make is more than SCOA by accumulating knowledge from the steganography and cryptosystems.

### 3.2    Attacks from Eve

There exists a unique equilibrium between two rational sides with certain prior knowledge and the fixed rules in game according to game theory. Based on the analysis of attack grading by Ke's [1], once the goal and the rules of the game are fixed, the maximum possible benefit of Eve is only associated with her prior knowledge [9][10]. According to Section II, stego-ciphertext can be received by legitimate receivers of the cryptosystem or eavesdroppers. Eve's ability is based on the above two identities. Attacks from Eve includes: stego-cover only attack (SCOA), known cover attack (KCA), chosen cover attack (CCA), adaptive chosen cover attack (ACCA), which are graded based on her prior knowledge of the systems of steganography and cryptography.

SIED involves two types of operations: encryption/ decryption, embedding/ extraction. The methodology of embedding is usually not a universal one, but proposed based on a specific encryption method. So, the cover includes not only the ciphertext, but also the plaintext and the encryption methods adopted in SIED. Intuitively, the more information about the targets obtained, the easier it is to achieve the attack, which is called a higher-level attack. The higher-level attack a algorithm can resist, the higher-level security it possesses.

### 3.3 Security of steganography in encrypted domain

**Security against SCOA**

SCOA is the primary level attack on SIED where the attacker is assumed to have access only to a set of stego-ciphertext. In addition, the attacker still has some knowledge of the natural ciphertext. The attacker, Eve, is an eavesdropper outside the cryptosystem and does not possess encryption or decryption keys. The pre-knowledge that Eve can use in SCOA is from the stego-ciphertext that exists in the channel, e.g., the basic size of ciphertext data, the complexity of decryption, and the theoretical distribution characteristics of ciphertext. For example, in reality, an attacker may monitor public channels to obtain the stego-ciphertext.

To ensure security against SCOA, an additional data expansion in the ciphertext should not be resulted by embedding, nor an increase in the decryption complexity. As discussed in Section III. B, the stego-ciphertext should follow the randomly uniform distribution, so that the Eq. (10) is satisfied to resist SCOA.

$$D(P(\boldsymbol{C}), P(\boldsymbol{C}'))=0 \text{ or } D(P(\boldsymbol{C}'), U(0, q))=0 \qquad (10)$$

**Security against KCA**

KCA is a steganalysis attack where, in addition to the prior knowledge under SCOA, Eve also has more knowledge from several pairs of the plaintext, the ciphertext, its corresponding stego-ciphertext and the decrypted result of the stego-ciphertext. The number of pairs is finite within the polynomial complexity. Eve can not only carry out the learning in SCOA, but also learn from the original ciphertext, the stego-ciphertext and the decrypted result of the stego-ciphertext. For example, in reality, in addition to monitoring all open channels, the attacker might be a legitimate receiver of the cryptosystem. She can normally decrypt the ciphertext. Compared with SCOA, Eve with KCA is more normalized and universal in applications.

Security against KCA requires the indistinguishability of the decrypted result of the stego-ciphertext and the plaintext besides the security requirements against SCOA.

$$\text{Dec}(\boldsymbol{K}_{\text{Enc}}, \boldsymbol{C}) = \text{Dec}(\boldsymbol{K}_{\text{Enc}}, \boldsymbol{C}') \qquad (11)$$

PSNR (Peak Signal to Noise Ratio) or MSE (Mean Square Error) can be introduced. If PSNR is ∞, it demonstrates that the Eq. (11) is satisfied.

**Security against CCA**

CCA：In addition to the prior knowledge under the KCA, the attacker can also invoke several times of embedding or extraction process of the current steganography system. The number of the invoking operation is finite within polynomial complexity. In addition to the learning phase of KCA, Eve can invoke the embedding or extraction process to learn from the changes of process variables in the processes of embedding or data extraction. The cover invoked can be a stego-cover or a forged one by Eve. For example, in reality, the attacker can research the systems of cryptography and SIED for several times, or instigate a user of the systems to return results operated step by step on the stego-ciphertext.

On the basis of all the requirements of security against KCA, security against CCA requires that all the process variables generated by each step of decrypting stego-ciphertext are indistinguishable from the variables generated by decrypting the original ciphertext. Let the set of process variables generated from decrypting the ciphertext $C$ be $\Theta=\{\delta_1, \delta_2, \delta_3, \ldots | \delta_i \leftarrow \text{Dec}(K_{\text{Enc}}, C)\}$, the set of process variables generated from decrypting the stego-ciphertext $C'$ be $\Theta'=\{\delta_1', \delta_2', \delta_3', \ldots | \delta_i' \leftarrow \text{Dec}(K_{\text{Enc}}, C')\}$. For any $i$,

$$D(\delta_i, \delta_i')=0 \quad \delta_i \in \Theta, \delta_i' \in \Theta' \tag{12}$$

**Security against ACCA**

In ACCA, if CCA fails, the attacker can restart the CCA learning phase on the targeted steganography system, and repeat the attacks several times within the polynomial complexity until she succeeds. For example, in reality, the attacker is in charge of the cryptography system.

On the basis of all the requirements of security against CCA, the security against ACCA requires to continuously resist Eve from accumulating useful experiences by repeating embedding and extraction processes. It requires that each operation in the embedding process should be a standard encryption operation. If there were inevitably nonstandard encryption operations for embedding, the distribution of each process variable should follow the same distribution as the one without embedding. Note that the distribution should be a simply modeled standard distribution, e.g., Gaussian distribution. If the distribution is too hard to model, it is difficult to effectively guarantee the indistinguishability of the distribution after embedding.

## 4  Realizations of SIED

Currently, it is mainly the technology of reversible data hiding in encrypted domain (RDH-ED) that supports data embedding in the ciphertext. However, not any a RDH-ED algorithm can act as the embedding function for SIED due to the limitation of security requirements. For example, the algorithm in [11] needs to expand the ciphertext for embedding. It cannot resist SCOA because of the change of the data size of stego-ciphertext after embedding. There are some other unsuitable RDH-DE algorithms for SIED, in which modifications were made to the encryption processes for embedding and the encryption strength and encryption parameters setting are compromised [12]. In this section, we have selected four typical RDH-ED algorithms as the embedding and extraction function of the realizations of SIED.

### 4.1  SIED scheme against SCOA

**Algorithm and Framework**

The algorithm of fully homomorphic encryption encapsulated difference expansion (FHEE-DE) [13] is adapted as the embedding function for SIED.

Since FHE in FHEE-DE only applies to ciphertext, the secret message must be encrypted before embedding and can only be extracted from the decrypted result by using DE extraction in spatial domain. In the data hider's side, Alice must encrypt the secret message by using the public key and then embeds the secret message into ciphertext by using FHEE-DE. Then the stego-ciphertext could be generated. In the extractor's side,

Bob has to use the secret key to decrypt the stego-ciphertext to obtain the marked plaintext. Secret message could be extracted from the marked plaintext. The framework is as shown in Fig. 5, which conforms to AC mode.

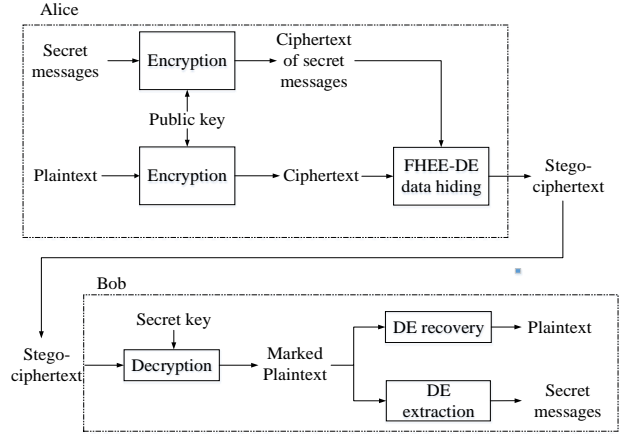

**Fig. 5.** The application framework of SIED based on FHEE-DE in 13

**Security analysis**

Since FHEE-DE is constructed based on standard FHE operations, the data distribution of stego-ciphertext is consistent with the normal ciphertext. There is no secondary data expansion resulted by embedding, thus Eq. (10) can be satisfied. Therefore, the scheme can resist SCOA.

However, the scheme based on FHEE-DE cannot resist KCA. As shown in Fig. 5, the secret message can only be extracted from the marked plaintext, so it is the marked plaintext instead of the original plaintext that is obtained after decrypting the stego-ciphertext. Though the distortion in the marked plaintext can be reduced to a low level (PSNR $\geq$ 65 dB in [13]) and a recovery process can be performed to obtain the lossless plaintext, it cannot prevent Eve who own the decryption key from finding the distortion in the decryption result to confirm the existence of secret messages.

### 4.2 SIED scheme against KCA

As analyzed in Section II, in practice, it is common to meet with Eve owning the decryption key. Therefore, a RDH-ED algorithm cannot act as the embedding function of SIED against KCA as long as the direct decryption distortion exists. To resist KCA is far more practical than to resisting SCOA.

**Algorithm and Framework**

The instance of SIED against KCA is based on Ke's algorithm in [14]. Controllable redundancy of LWE cryptosystem is utilized for embedding in encrypted domain. Scheme [14] takes advantage of the redundancy of quantization element to embed data and there is no additional data expansion resulted by embedding and the encryption strength is well maintained. The framework is as shown in Fig. 6. Alice embeds the secret message into the ciphertext to obtain the stego-ciphertext. Then it is transported to Bob, who has the secret key to extract secret messages from stego-ciphertext. It should be noted that the directly decrypted result of stego-ciphertext is the plaintext and no recovery process is needed. This mode conforms to EXC mode.

**Security analysis**

The embedding processes of [14] would not result in any secondary data expansion of ciphertext. It has been deduced in 14 that the stego-ciphertext follows the same distribution as original ciphertext, thus meeting the security requirement of D($P(C)$, $P(C')$)=0. According to the theoretical analysis of the extraction function and the experimental results in [14], the directly decrypted result of the stego-ciphertext is the lossless plaintext, which meets the requirement in Eq. (11): Dec($K_{Enc}$, $C$) = Dec($K_{Enc}$, $C'$). Therefore, the SIED based on [14] could resist KCA.

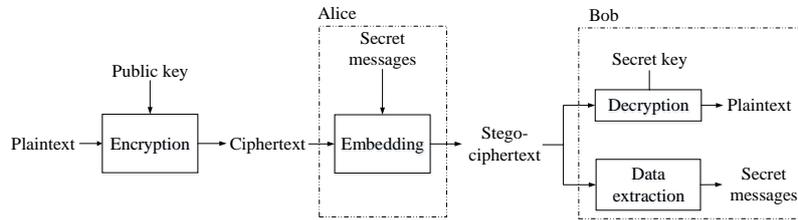

**Fig. 6.** The application framework of SIED based on RDH-ED in [14].

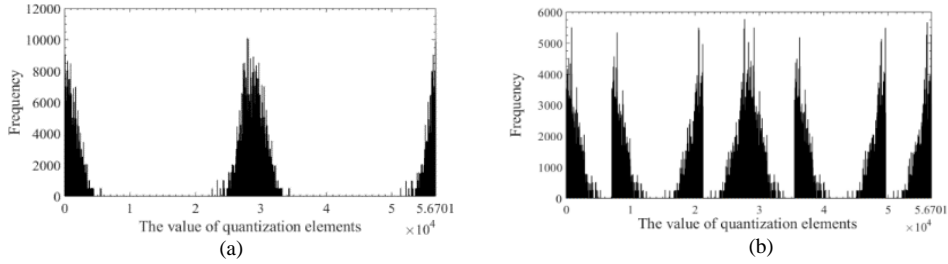

**Fig. 7.** Distributions of quantization vectors from (a) Original ciphertext; (b) Stego-ciphertext.

To resist CCA, all the possible open encryption parameters after embedding should not reveal any clues to the existence of steganography. However, the quantization vectors from original ciphertext and stego-ciphertext follow the different distributions as shown in Fig. 7.

There are more peaks in Fig. 7(b) than in Fig. 7(a). Namely, D($\delta_i$, $\delta_i'$)≠0 $\delta_i \in \Theta$, $\delta_i' \in \Theta'$. Eve can learn the differences to confirm the presence of the steganography with a high probability. It cannot resist CCA.

### 4.3 SIED scheme against CCA

**Algorithm and Framework**

In this section, a lossless data hiding method in encrypted domain based on encryption variable refreshing (EVR) algorithm was proposed, which could act as the embedding function of the realization of SIED against CCA. EVR is based on the cryptography with randomly sampled variables in the encryption process, such as Paillier encryption. By refreshing the variable randomly, another ciphertext is generated. The bits of ciphertext are sampled from specific positions. If the sampled bits match the to-be-embedded bits, the stego-ciphertext is obtained. If the bits are not matched, the variable is refreshed again until the resampled bits are matched. The embedding function of EVR for SIED based on Paillier encryption is as following.

*Key generation*

Choose two big prime $p$ and $q$, and calculate $N=p \times q$. Calculate the lowest common multiple of ($p$-1, $q$-1) denoted as $\lambda$. Choose an integer $g \in Z^*_{N^2}$, gcd(L($g^\lambda \bmod N^2$), $N$)=1, L($x$)=($x$-1)/$N$, $Z^*_{N^2}$ is the set of integers that are less than $N^2$ and relatively prime to $N^2$. The public key for encryption is $K_1(N, g)$, the secret key for decryption is $K_2(p, q, \lambda)$.

*Encryption*

Choose an encryption process variable randomly $r \in Z^*_N$. The plaintext is $p \in Z_N$, the encrypted result is $c = \text{Enc}(K_1, p) = g^p \cdot r^N \bmod N^2$.

*Decryption*

The decrypted result of $c$ is $p = \text{Dec}(K_2, c) = \dfrac{L(c^\lambda \bmod N^2)}{L(g^\lambda \bmod N^2)} \bmod N$.

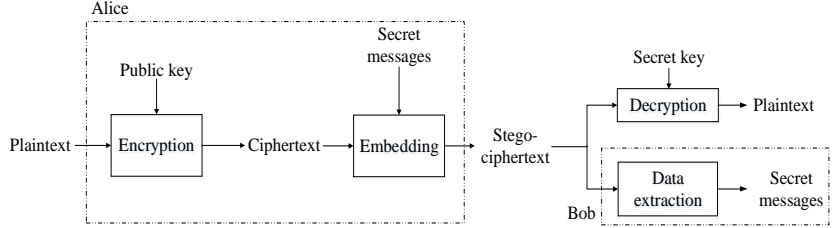

**Fig. 8.** The application framework of SIED based on EVR.

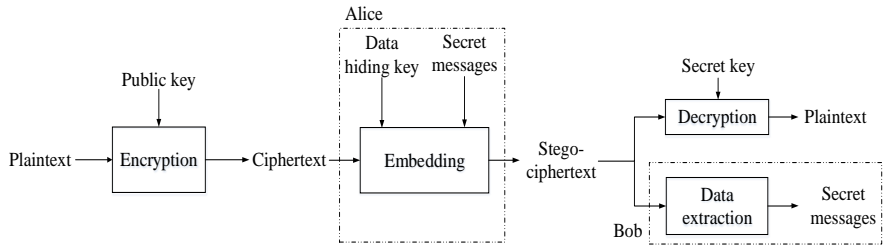

**Fig. 9.** The application framework of SIED based on KS-LSB.

*Bit sampling embedding by EVR*

The to-be-refreshing variable is $r$. The to-be-embedded bit is denoted as $b_s$. LSB(.) is to obtain the least significant bit of the input integer.

Step1: If $b_r = \text{LSB}(c)$, $c' = c$; if $b_r \neq \text{LSB}(c)$, *refresh* $r' \neq r$.

Step 2: $c' = \text{Enc}(K_1, p) = g^p \cdot r'^N \mod N^2$.

Step 3: Repeat Step 2 until $\text{LSB}(c') = b_s$. Return $c'$ as the stego-ciphertext.

Decryption: $m = \text{Dec}(K_2, c')$.

Extraction: $b_r = \text{LSB}(c')$.

The application is shown in Fig. 8. Alice needs to refresh the variables to obtain the stego-ciphertext. Bob can eavesdrop on the stego-ciphertext from the transmission channel. The decryption key is not necessary for Bob, even if he is a legitimate receiver of the cryptosystem. The decryption and extraction processes are independent. Therefore, the plaintext can be obtained without loss by directly decrypting the stego-ciphertext. It conforms to EMC mode.

**Security analysis**

In CCA, Eve cannot obtain any useful information from the stego-ciphertext which is another normal ciphertext, namely, $D(P(C), P(C')) = 0$. No ciphertext expansion exists and the computational complexity of decryption is not increased. There is no decryption distortion, *i.e.*, $\text{Dec}(K_{\text{Enc}}, C) = \text{Dec}(K_{\text{Enc}}, C')$. Since the operations of EVR embedding are all standard encryption operations, all the process variables conform to the distribution of the original ones of Paillier encryption, that is, it satisfy Eq. (12): $D(\delta_i, \delta_i') = 0$ $\delta_i \in \Theta$, $\delta_i' \in \Theta'$. Therefore, the SIED based on EVR could resist CCA.

In ACCA, Eve could repeat learning from the generation of stego-ciphertext. Since Paillier encryption cannot resist the adaptive chosen ciphertext attack [15] and embedding process are part of the encryption process, Eve could learn from the adopted Paillier encryption to know about the bit sampling by EVR. Therefore, there is a high probability that the repeated sampling process will be suspected.

To resist Eve capable of ACCA, there might be two solutions for consideration. One is to adopt the cryptosystem that resists adaptive chosen ciphertext attack. It puts forward higher requirements for the encryption environment. The second is to construct the embedding method in which the embedding process and the encryption process are separable.

### 4.4 SIED scheme against ACCA

**Algorithm and Framework**

We introduce key-switching based LSB (KS-LSB) algorithm in [13] as the embedding function of the realization of SIED against ACCA. In the embedding process of KS-LSB, the data-hiding key is constructed from the switching key, which is independent of the encryption key and can be openly published. The extraction is independent of the decryption process. Bob can extract data without using the decryption key. Therefore, it conforms to the AF mode.

The application is shown in Fig. 9. Alice implements KS-LSB to obtain stego-ciphertext without using the encryption key. It is then transmitted to the receiver of the cryptosystem. Bob can eavesdrop on the stego-ciphertext and extract data from the stego-ciphertext without the decryption key. By directly decrypting the stego-ciphertext, the plaintext can be obtained without loss, and the decryption and extraction processes are independent of each other.

**Security analysis**

No data expansion of ciphertext or additional computational complexity of decryption is resulted by embedding. The directly decrypted result of stego-ciphertext is the lossless plaintext, namely, Dec($K_{Enc}$, $C$) = Dec($K_{Enc}$, $C'$). KS-LSB generates another standard ciphertext and Eve cannot gain any useful information from the stego-ciphertext or the process variables, namely, D($P(C)$, $P(C')$)=0 and D($\delta_i$, $\delta_i'$)=0 $\delta_i \in \Theta$, $\delta_i' \in \Theta'$. Key-switching is implemented in lattice-based encryption that is proved secure and could resist adaptive chosen ciphertext attack [13]. It could well encapsulate KS-LSB embedding, so that Eve cannot learn any difference existing in the variables before and after embedding by repeating analyzing the embedding or extraction operations. And the processes of embedding and encryption are separable. Therefore, the SIED based on KS-LSB could resist ACCA.

## 5      Conclusion

With the popularization and application of privacy protection technologies in cloud service and social network, ciphertext has been gradually becoming a common platform for public to exchange data. In this paper, we propose the steganography in encrypted domain. Based on Simmons' model of prisoners' problems, we discuss the application scenarios and security requirements of SIED. According to the function of the encryption key and decryption key in data embedding and extraction, the application modes of SIED are classified into four modes. Four levels of steganalysis attacks are introduced based on the prior knowledge about the steganography system that the attacker is assumed to obtain. And then the four levels of steganography security of SIED are defined correspondingly. Based on the existing reversible data hiding algorithms, we give four schemes of SIED as practical instances with different security levels. By analyzing the embedding and extraction characteristics of each instance, the SIED modes, application frameworks and security levels are discussed in detail.

## Acknowledgment

This work was supported by National Natural Science Foundation of China under Grant No.62102450, No. 61872384 and No. 62272478 fundamental research fund project of Engineering University of PAP under Grant No. WJY202112. The authors also gratefully acknowledge the helpful comments and suggestions of the reviewers.